\RequirePackage{fix-cm}
\documentclass[smallextended]{svjour3}       
\smartqed  
\usepackage{graphicx}
\usepackage{color}
\usepackage{amsmath}
\makeatletter

\newcommand{\Rmnum}[1]{\expandafter\@slowromancap\romannumeral #1@}
\makeatother

\begin{document}

\title{Dynamics of a system of coupled inverted pendula with vertical forcing
}


\author{Nivedita Bhadra   
}


\institute{ \at
              Department of Physical Sciences \\
              Indian Institute of Science Education \& Research, Kolkata\\
              Mohanpur Campus, West Bengal--741246, India\\
              \email{nivedita.home@gmail.com}
}

\date{Received: date / Accepted: date}

\maketitle

\begin{abstract}
Dynamical stabilization of an inverted pendulum through vertical movement of the pivot is a well-known  counter intuitive phenomenon in classical mechanics. This system is also known as Kapitza pendulum and the stability can be explained with the aid of effective potential. We explore the effect of many body interaction for such a system. Our numerical analysis shows that interaction between pendula generally degrades the dynamical stability of each pendulum. This effect is more pronounced in nearest neighbour coupling than all-to-all coupling and stability improves with the increase of the system size. We report development of beats and clustering in network of coupled pendula.
\end{abstract}

\section{Introduction}
\label{intro}
An ordinary rigid planar pendulum with a fixed suspension point has
only one stable configuration, the bob hanging below the suspension
point. However, if a vertical periodic force is applied at the
suspension point, the system becomes stable in the vertical position
also, provided the amplitude and the frequency are kept within 
certain intervals. This is an example of dynamic stabilization and has
several applications in the field of atomic
physics \cite{Paul1990,Gilary2003}, plasma physics \cite{Bullo2002} and
cybernetics \cite{Nakamura1997}. This unusual and counterintuitive
phenomenon was first experimentally observed by Stephenson 
\cite{Stephenson1908} and theoretically explained by P. Kapitza 
\cite{Haar(Eds.)1965}. The system is also known as Kapitza
pendulum. There exists a large volume of literature in this field,
both experimental and theoretical
\cite{Kim1998,Butikov2001,Broer1999,Broer2004,Bartuccelli2001,Bartuccelli2002,Bartuccelli2003}.
The potential function is time dependent which leads to dynamic stabilization.\par
What happens when such systems interact with each other? What is the effect of coupling in such a system with spatio-temporal potential? Of late, the understanding of periodically driven systems has become one of the most active areas of research in many body physics. 
 In the context of spatiotemporal chaos, coupled Kapitza pendulum in a dissipative environment has been investigated in \cite{Chacon2010}. Ref.\cite{Marcheggiani2014} addresses the issue of synchronization in a coupled Kapitza pendulum where the coupling is mechanically linear and geometrically nonlinear. Parametric oscillator in several contexts has been rigorously investigated in previous works  \cite{Moehlis2008,Danzl2010,Xu2014,SalgadoSanchez2016}.\par
In this work, we specifically address the effect on the stability range around the inverted position due to the presence of interaction between pendula. A recent work by Citro et al. \cite{Citro2015} addressed this question in a network of pendula with nearest neighbour coupling. The coupled Kapitza pendula in the continuum limit is shown to take the form of a
periodically-driven Sine-Gordon model. If interaction is
introduced, the stabilization property changes which is characterised by the interaction strength. In recent times, the study of periodically
driven many body systems has drawn attention due to the recent
advancement in the field of ultra-cold
atoms \cite{Eckardt2005,Lignier2007,Struck2011,Struck2012}.

It is often assumed that a many-body system undergoing a cyclic
process will reach infinite temperature as demanded by the second law
of thermodynamics. But several research works have shown that periodic
heating and cooling is possible in such a system 
\cite{Russomanno2012,Ponte2015,Ponte2015a}. Apperance of parametric
resonance can explain this phenomenon by separating the absorbing and
non-absorbing regimes. Such interactions have so far been considered only in a 1-D chain. Do the same result hold in a 2D lattice? What is the effect of system size(the number of interacting pendula)?\par
The effect of long range interaction has
not been addressed in these works. Moreover, the effect of nearest
neighbor coupling has been studied only in a 1D chain. We study the effect of all to all interaction in our present work and
compare with the result of the 1D case. We have investigated the
effect of the nearest neighbour interaction in a 2D network as well.

The dynamics of the particle system with all to all interaction i.e.,
long range interaction offers several unusual and new characteristics
in comparison with systems with short-range or nearest neighbour
interactions. The Kuramoto oscillator and the Hamiltonian mean field model
are two iconic test-beds for such class of systems. This kind of long
range interaction demands statistical treatment. 
The state of the art method to deal with this class of problems is the
mean field theory (MFT), where all the interactions are considered to
be of equal strength. Through the mean field treatment the complexity
of the system reduces drastically. In this work, we use numerical simulations to study the coupled pendulum system with all to all
interaction. Our attempt is to find the stability
regime for the system and dynamics of the coupled system. We numerically obtain the maximum
release angle beyond which the stabilization of the inverted state is
not achievable.

 We also obtain how this maximum release angle, which is a measure of
the stability of the system, changes with the size of the system $N$
and the type of coupling. We consider three different cases for the
coupled system. Our study includes nearest neighbour interaction as well as all
to all interaction.  To observe the effect of dimension we investigate
a 2D square network. Finite size effect, i.e., dependence of stability on
the total number of pendula has also been discussed in this
context.  

The paper is organized as follows. In
Sec. \ref{sec:SingleKapitza} we briefly recall the single
Kapitza pendulum and its stability condition. In Sec. \ref{sec:MBK} we consider many body interaction of different types. We present some concluding remarks in Sec. \ref{sec:Discussion}.
\section{Single Kapitza pendulum}\label{sec:SingleKapitza}

\begin{figure}[h!]
\centering
\includegraphics[scale=.65]{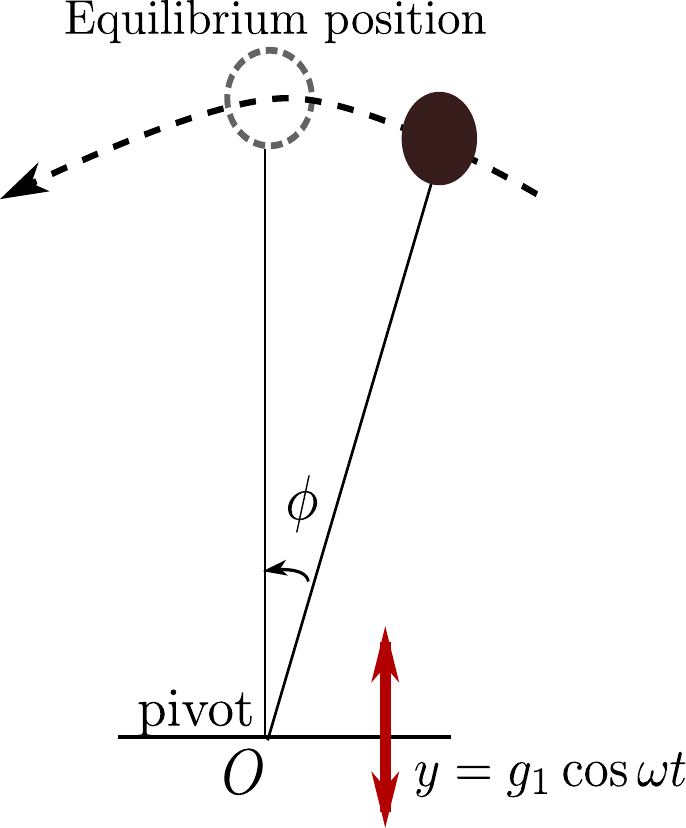}
\caption{Schematic diagram of single, isolated Kapitza pendulum. O is the point of suspension subject to a high frequency vertical drive $y=g_1\cos{\omega t}$. $\phi$ represents the angular dispacement.}
\label{fig:singlekp}
\end{figure}

Here we briefly summarize the case of a single inverted pendulum where the pivot oscillates vertically with amplitude $g_1$ and frequency $\omega$ (Fig.\ref{fig:singlekp}). The Hamiltonian governing the system is
\begin{eqnarray}
\label{eq:SingleKP}
H=\frac{p^2}{2}-g(t)\cos{\phi},
\end{eqnarray}
 where $p$ and $\phi$ are momentum and angular displacement of the pendulum respectively and $g(t)=g_0+g_1 \cos{\omega t}$. When the
vertical drive is absent i.e., $g_1=0$, there are two fixed points $\phi=0$ (stable)
and $\phi=\pi$ (unstable). If $g_1\neq 0$, the inverted position
i.e., $\phi=\pi$ configuration can be stable if $\omega$ is very high
compared to the natural frequency $\sqrt{g_0}$ . In order to understand the
stable configuration for this time dependent potential system an
effective potential is required to be calculated which was for the
first time obtained by Kapitza by separating the `fast' and `slow'
variables and integrating out the `fast' ones. The condition for
stabilization of the inverted position turned out to be $g_1^2\geq
g_0 \omega^2/2$. 
 This concept of effective potential gave birth to a
new branch of physics, named dynamical stabilization. 
The method to obtain the effective potential for such a system is perturbative and the result is approximate. One can obtain the exact trajectory only through numerical analysis.
\begin{figure}[t]
\includegraphics[scale=.4]{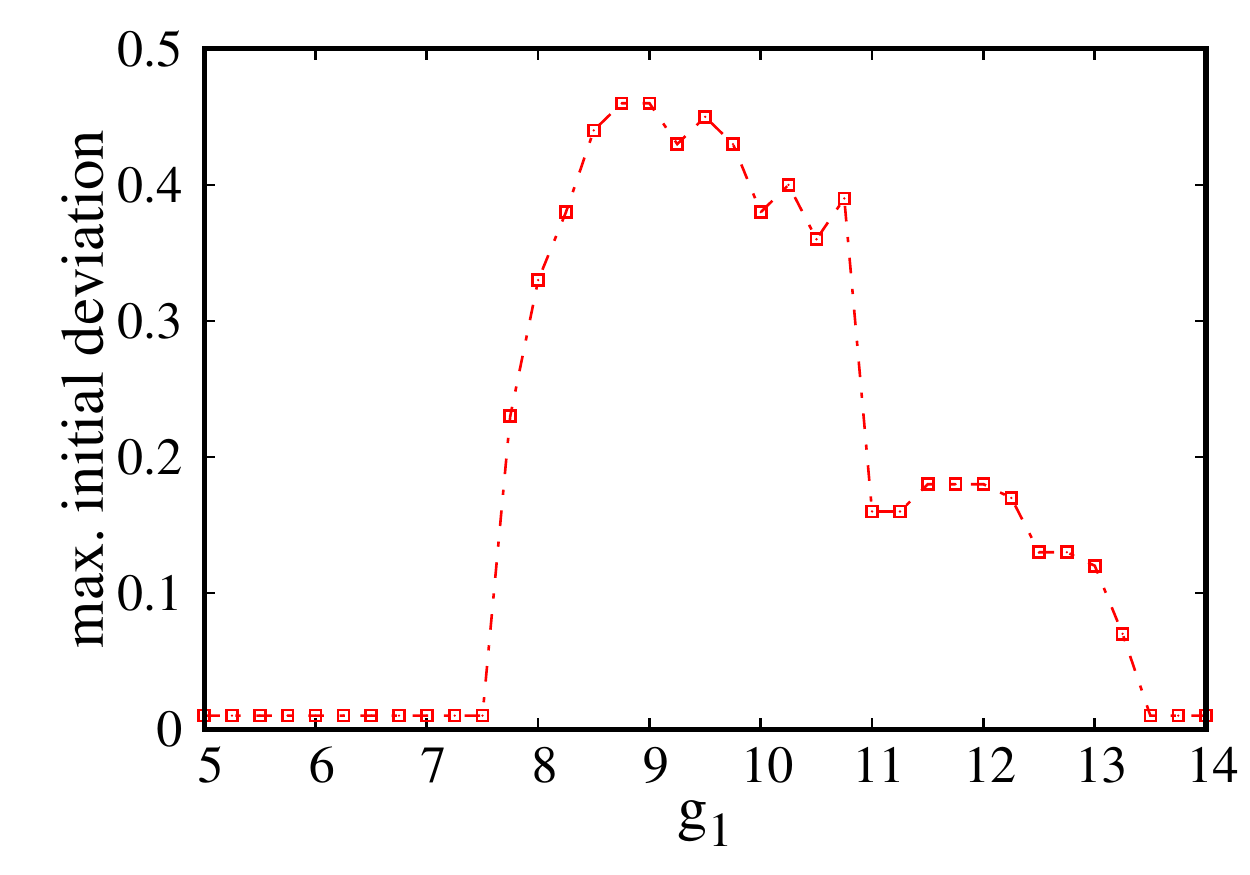}(a) \hspace{.2cm}
\includegraphics[scale=.4]{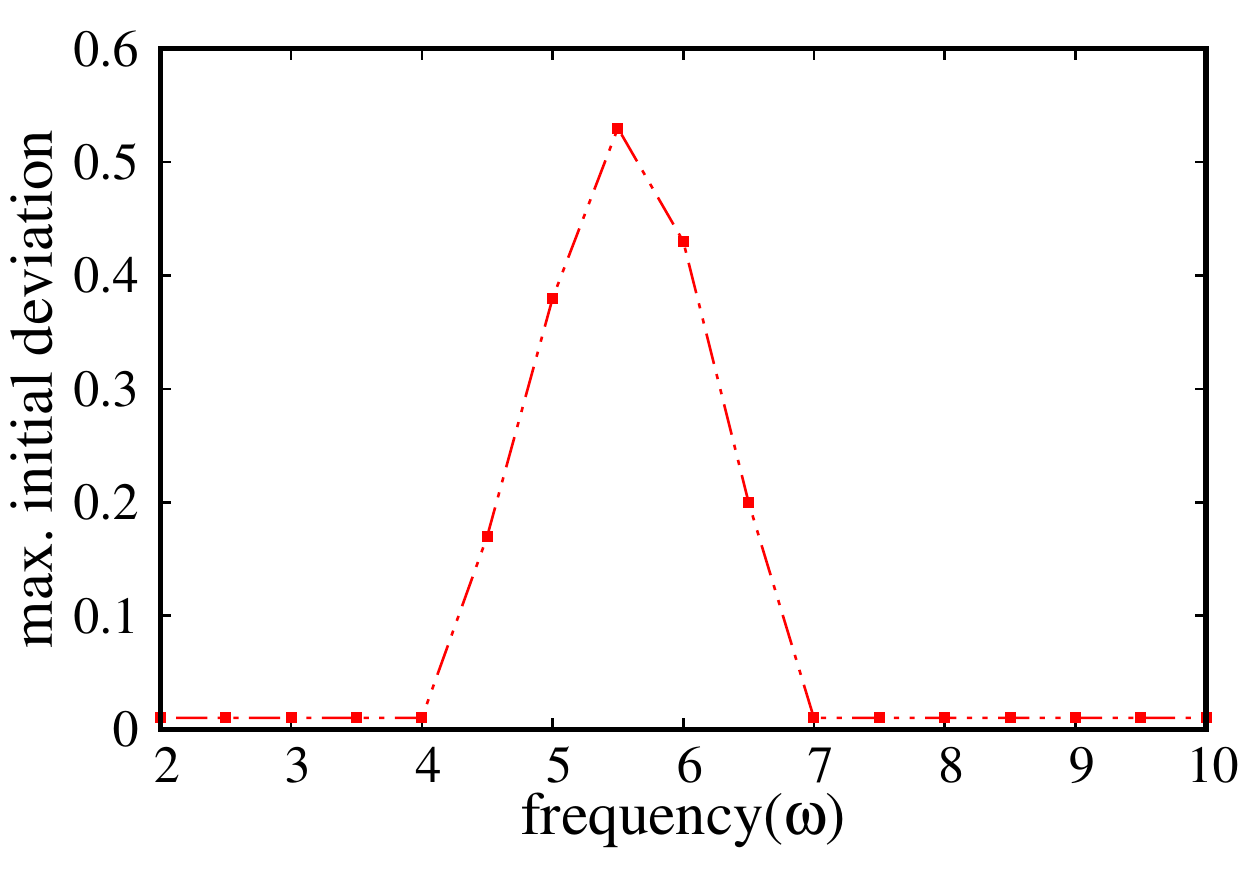}(b)
\caption{Stability of a single Kapitza pendulum: the maximum initial deviation from the inverted position $\phi=\pi$ for stability (a) for different values of applied amplitude $g_1$, with $g_0=1, \omega=5$, (b) for different values of applied frequency $\omega$ with $g_0=1, g_1=10$. }
\label{fig:SingleKPamp}
\end{figure}

From the effective potential, we obtain the shape of the potential
well. It can be shown that if the driving frequency is fixed at some
value and the amplitude of the drive is increased, the well becomes
wider \cite{Blackburn1992}. The maximum initial displacement for stabilization of the inverted pendulum quantifies the extent of dynamical stabilization for the system.
 We compute the maximum release angle beyond which the stabilization at the inverted position is lost for the system. This is a measure of the
 stability region around the inverted position of this
 system. Fig.\ref{fig:SingleKPamp}(a) shows how the maximum initial
 angle changes as we change the parameter $g_1$. This probes the width
 of the potential well for the system. Fig.\ref{fig:SingleKPamp}(b)
 represents the stability regime as the frequency of the drive is
 changed. These two figures show that, in case of the single Kapitza
 pendulum, there exists a range of initial deviation from the vertical
 position for which the inverted position is stable, and that this
 range is maximum at specific values of $g_1$ and $\omega$.

To compute the numerical solutions a $4^{\rm th}$ order Runge Kutta method
was employed, with time step $0.01$ times the period of the drive
($\frac{2\pi}{\omega}$). Now we explore how these ranges are affected when multiple Kapitza pendula interact with each other.

 \section{Many body Kapitza pendula}\label{sec:MBK}

\begin{figure}[t]
\centering
\includegraphics[width=0.75\textwidth]{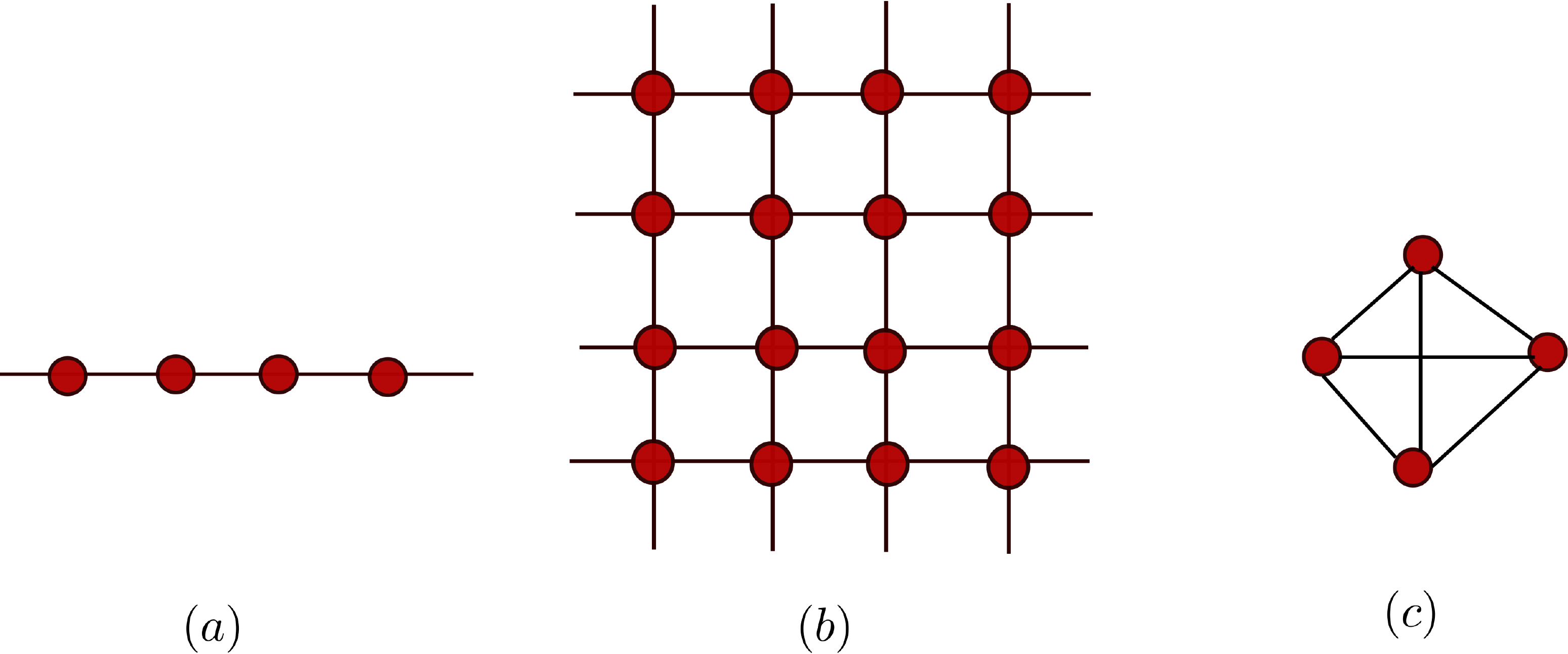} 
\caption{Schematic diagram of the coupled inverted pendulum. (a) 1D pendulum chain with nn uniform coupling (b) 2D pendulum network with nn coupling (c) Pendulum network with all to all coupling.}
\label{fig:pendulumchain}
\end{figure}

Now we study the stability of Kapitza pendula when more than one body
is involved. We observe several departures from the single particle
dynamics of this system. In this work, we explore three cases:
\begin{enumerate}
\item nearest neighbour
(nn) interaction in a 1D pendulum chain, \item 2D pendulum chain with
nn interaction, and \item coupled system with all-to-all
interaction. 
\end{enumerate}
We assume uniform coupling in all these three
cases. These three cases can be demonstrated in the form of three
different lattice structures where each lattice point represents the
bob of an inverted pendulum (Fig.~\ref{fig:pendulumchain}). For
example, in the square lattice of Fig.~\ref{fig:pendulumchain}(b),
each point is the pendulum bob and each one is connected to its
nearest neighbour, i.e., every pendulum is connected to its four neighbour
pendula.

 To observe
the effect of this interaction between the pendula we introduce a parameter $\Lambda$
into the Hamiltonian, which is a measure of
relative strength of the interaction between the pendula and the
force acting on individual pendula. 

\subsection{Case 1: 1D chain with nearest neighbour (nn) interaction}

\begin{figure*}[t]
\begin{center}
\includegraphics[width=.45\textwidth]{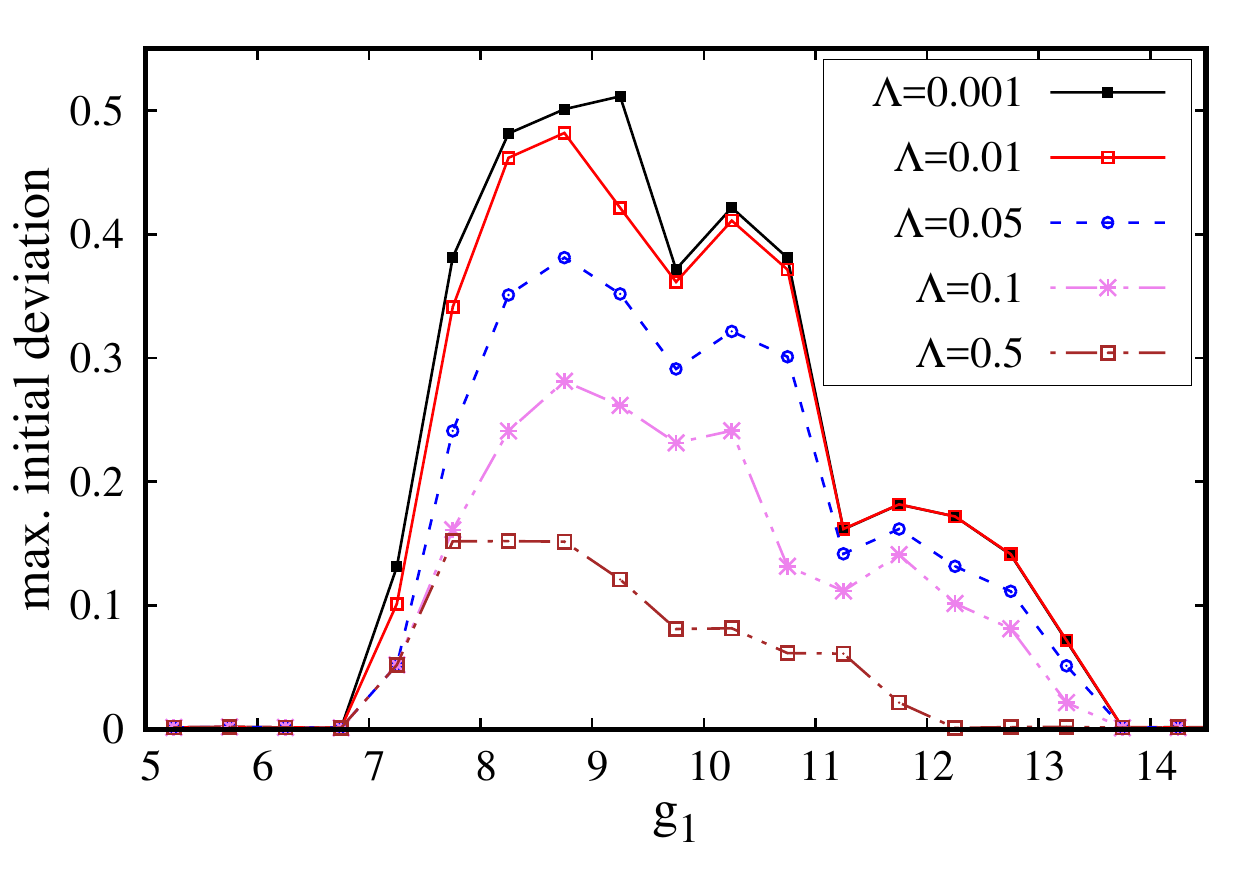}{\footnotesize (a)}\hfill 
\includegraphics[width=.45\textwidth]{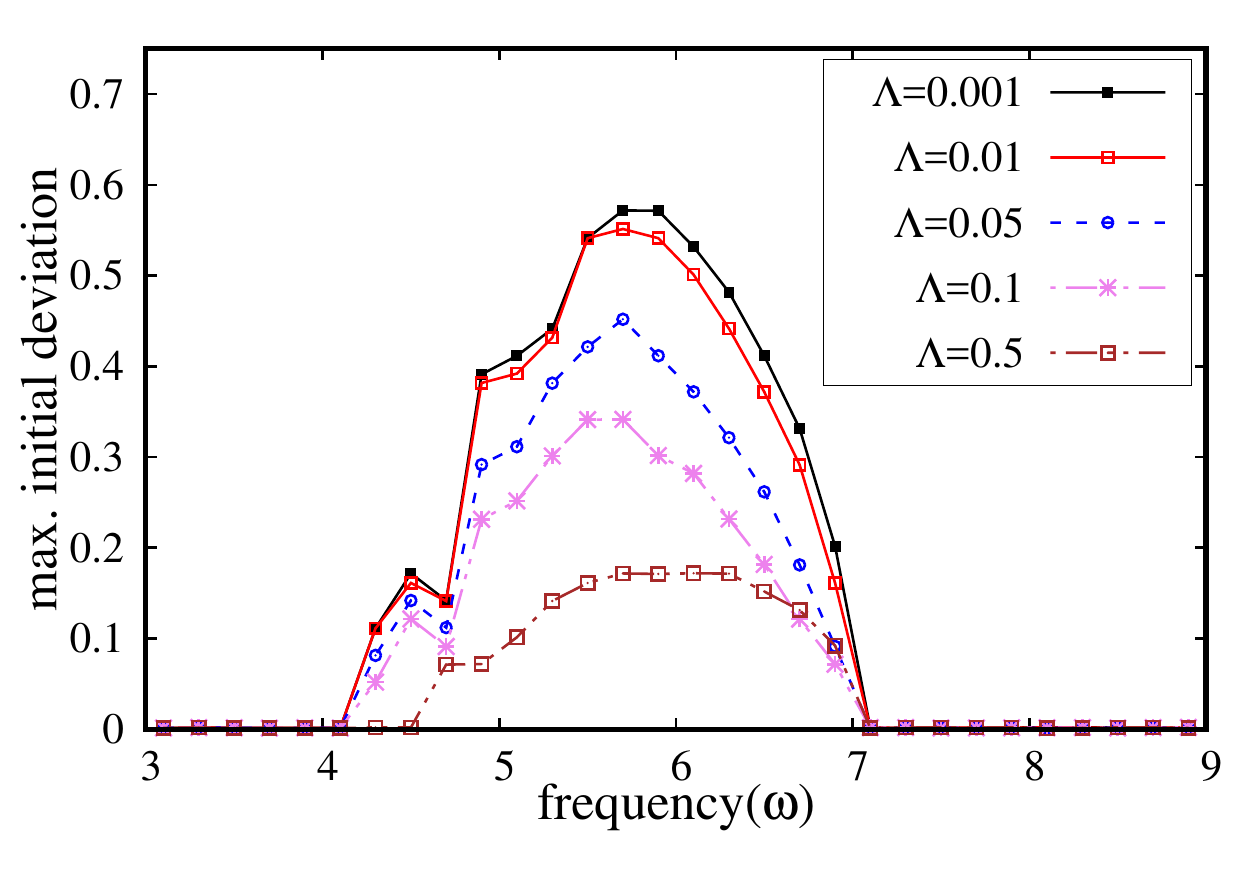} {\footnotesize (b)}\\ \medskip
\includegraphics[width=.45\textwidth]{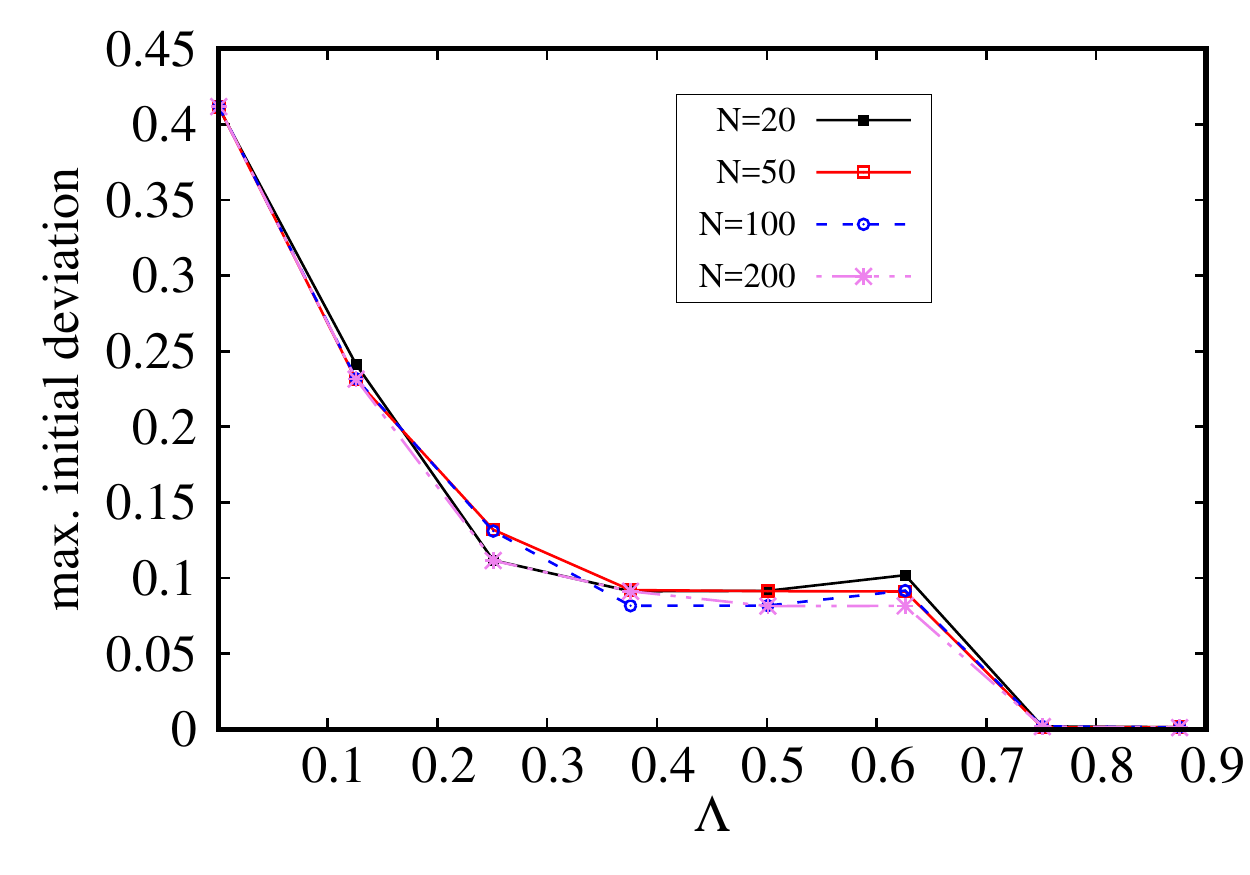} (c)
\end{center}
\caption{Stability of a 1D chain of Kapitza pendula with nearest neighbour coupling: the maximum initial deviation from the inverted position $\phi=\pi$, (a) for different values of applied amplitude for a set of coupling terms $\Lambda$. The other parameters are $g_0=1.0, \omega=5, N=100$. (b) for variation of the applied frequency for a set of the parameter $\Lambda$. The other parameters are $g_0=1.0, g_1=10, N=100$. (c) for variation of $\Lambda$ for different sizes of the system. The other parameters are $g_0=1.0, g_1=10,  \omega=5$.}
\label{fig:PD_nn}
\end{figure*}

We consider a system of $N$ coupled identical Kapitza pendula, $N$ being large. Following Ref. \cite{Citro2015} the system can be cast into the following Hamiltonian
\begin{eqnarray}
\label{eq:mbk_nn}
H= \sum_{i,j=i\pm 1}\Lambda\Big(\frac{ p_i^2}{2}-\cos(\phi_i-\phi_{j})\Big)-\sum_{i}\frac{g(t)}{\Lambda}\cos{\phi_i}.
\end{eqnarray}
 The energy scale $\Lambda$ determines the relative importance of the
 coupling between the pendula and the potential energy of individual
 pendulum. The coupling is restricted to the nearest neighbours only. We assume periodic boundary condition. The system equations are
 \begin{subequations}
  \begin{align}
\dot{\phi_i}&=\Lambda p_i, \\
\dot{p_i}&=-\Big( \frac{g(t)}{\Lambda}\sin{\phi_i}+\Lambda\sin(\phi_i-\phi_j)\Big).
 \end{align}
 \end{subequations}

\begin{figure}
\centering
\includegraphics[width=1\textwidth]{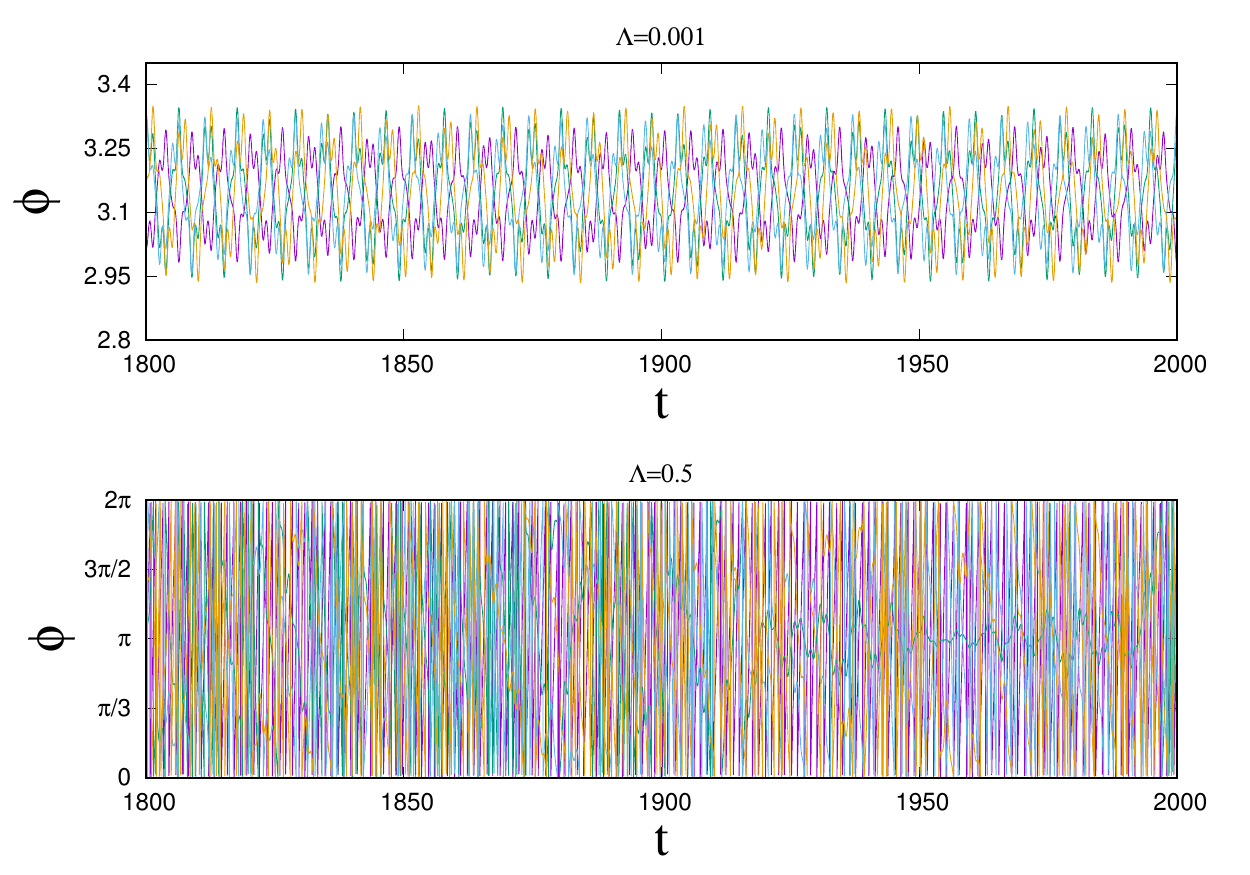}
\caption{(Color online) Time series for nn coupling for different values of $\Lambda=0.001$ (upper panel). We observe only oscillation of $\phi$ around the inverted position. For $\Lambda=0.5$, all pendula start to rotate i.e., stability is completely lost(lower panel). Plots are shown after the transients die down (here, we show time-series for $t=1800$ to $t=2000$).}
\label{fig:nnts}
\end{figure}

\begin{figure}
\centering
\includegraphics[width=.75\textwidth]{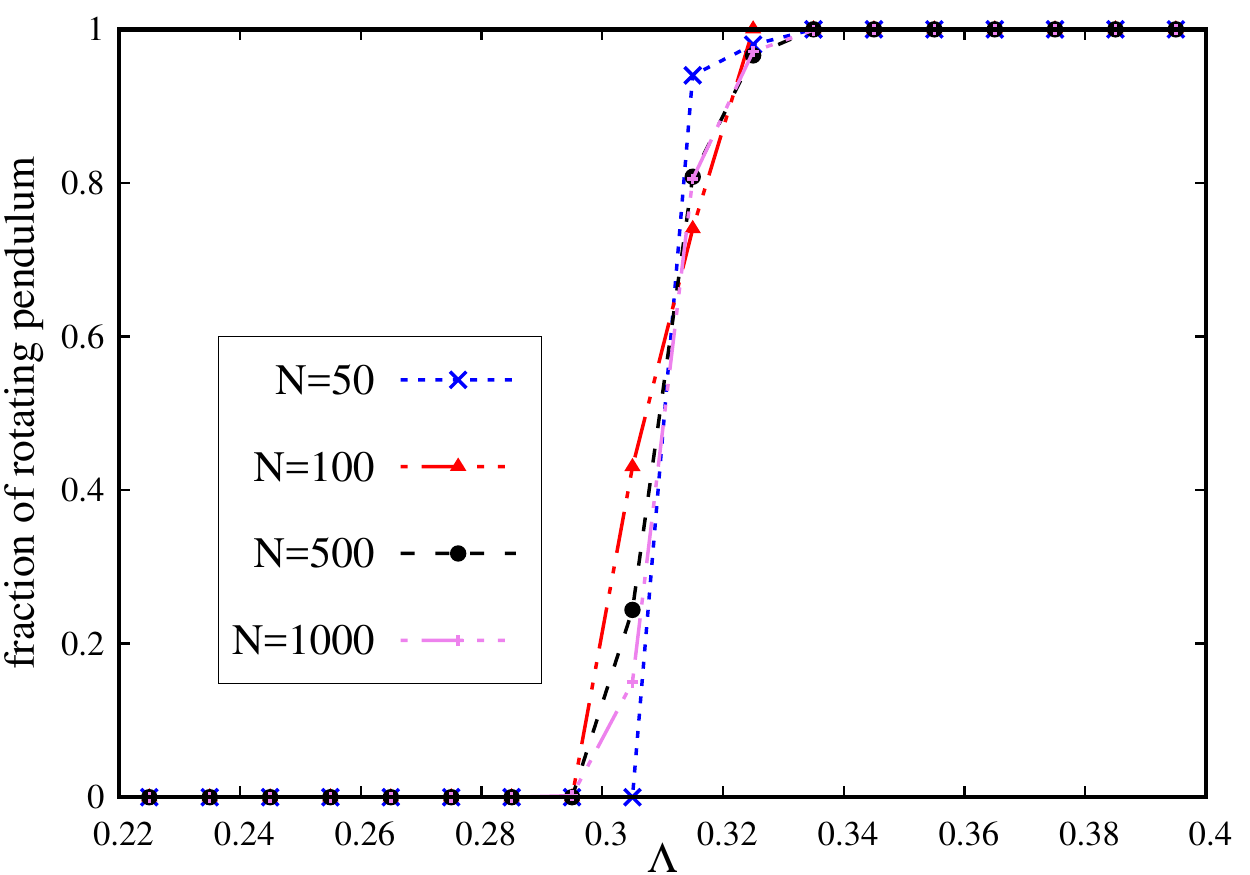}
\caption{(Color online) Fraction of rotating pendula for nearest neighbour coupling for different system sizes. For small values of $\Lambda$ each pendulum shows oscillation around the inverted position, For a very small range of $\Lambda$ the system shows both oscillation and rotational motion. After crossing the threshold every pendulum starts to rotate.}
\label{fig:fracrotnn}
\end{figure}
The numerical results are presented in Fig.~\ref{fig:PD_nn}.
Fig.~\ref{fig:PD_nn}(a) shows the maximum initial angle versus $g_1$
curves.  As we increase the value of $\Lambda$ the height of the curve
decreases implying decrement of stability range. As the interaction
between the nearest neighbours increases gradually, stabilization is
disturbed. In the simulation, each pendulum is started from slightly different initial conditions. The initial angular position($\phi$) of each pendulum is taken
to be $\pi+0.001$ with some random deviation($\mathcal{O}(10^{-3})$)
for each pendulum.\par
 Fig.~\ref{fig:PD_nn}(b) shows the curves for maximum
initial angle versus the driving frequency. It exhibits a similar
nature of the shape of the curves. It shows that the system is stable only over a range of $\omega\sim [4,7]$ and that the stability margin degrades with increase of the coupling strength $\Lambda$. Fig.~\ref{fig:PD_nn}(c) shows the
effect of the size of the system. We have studied cases for $N=20, 50,
100, 200, 500$.  It shows that the stability reduces with increase in
$\Lambda$, and that this result does not significantly depend on the
system size $N$. 

Now we explore the coupled behaviour in the system. The time series for
this system is shown in Fig.\ref{fig:nnts}. It shows that for low
values of $\Lambda$ (e.g., $\Lambda = 0.001$), each
pendulum oscillates (stable mode) around the inverted position. This behavior is lost when $\Lambda$ is increased to
around $0.5$. Most of the pendula lose stability (go into rotating motion). We show the fraction of pendula going into rotational motion as we increase the interaction $\Lambda$ in Fig. \ref{fig:fracrotnn}. We find that for different system size, the rotational motion starts when $\Lambda\approx 0.3$. Each point in the plot is averged value of $20$ different initial conditions for each pendulum. 

\subsection{Case 2: 2D square network of Kapitza pendula with nearest neighbour coupling}

 In order to understand the effect of dimension we investigate the case of a 2D
 square network with nn coupling. Following the same procedure as in the 1D case, we construct the Hamiltonian for this case which can be
written as \begin{eqnarray}
\label{eq:mbk_2D}
H= \sum_{i,j,k,l}\Lambda\Big(\frac{ p_{i}^2}{2}-\cos(\phi_{i,j}-\phi_{k,l})\Big)-\sum_{i,j}\frac{g(t)}{\Lambda}\cos{\phi_{i,j}},
\end{eqnarray}
where $i,j$ represent indices in $x$ and $y$ directions and $k=i\pm 1, l=j\pm 1$. Here the
interaction is with the nearest neighbours only, i.e., each pendulum
is connected to its four nearest neighbours. 
\begin{figure*}[t]
\begin{center}
\includegraphics[width=.45\textwidth]{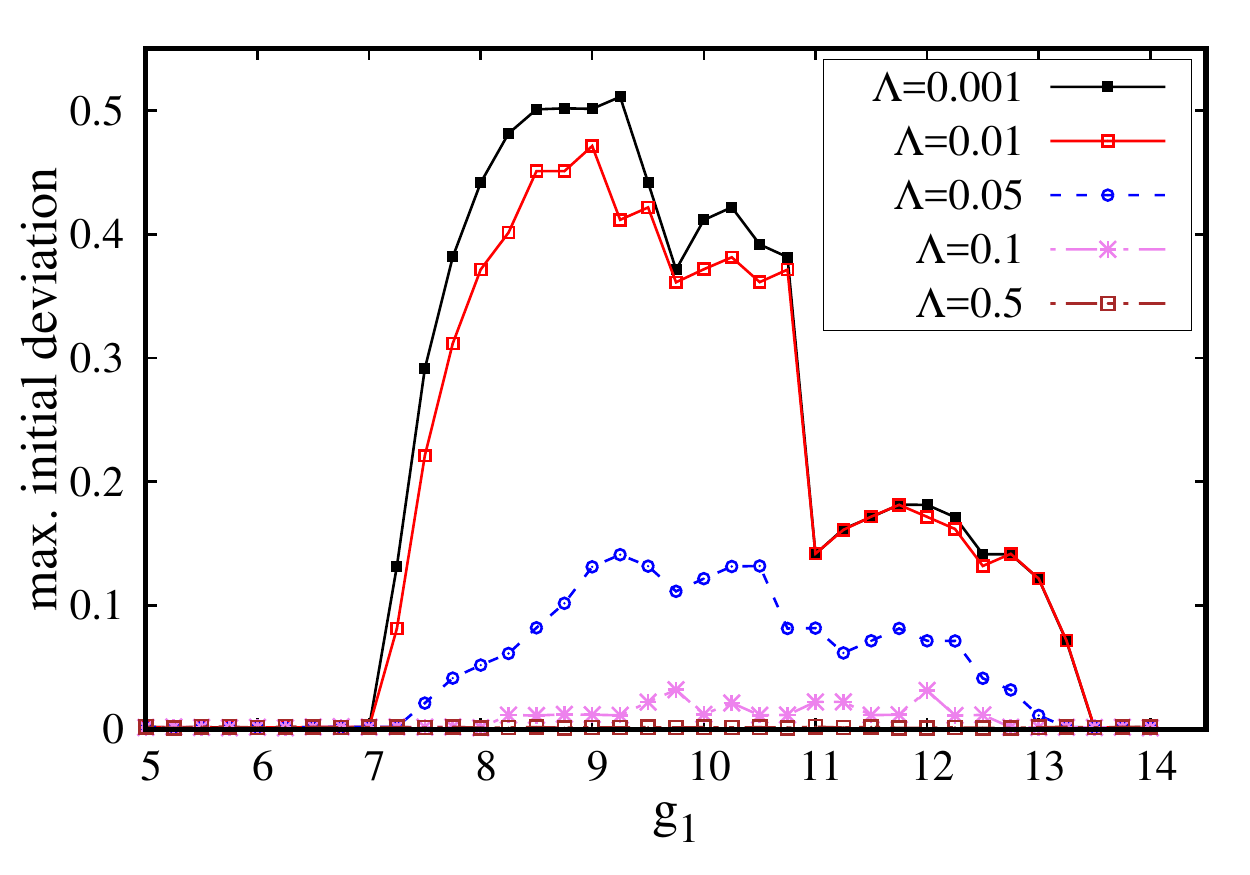}{\footnotesize (a)}\hfill
\includegraphics[width=.45\textwidth]{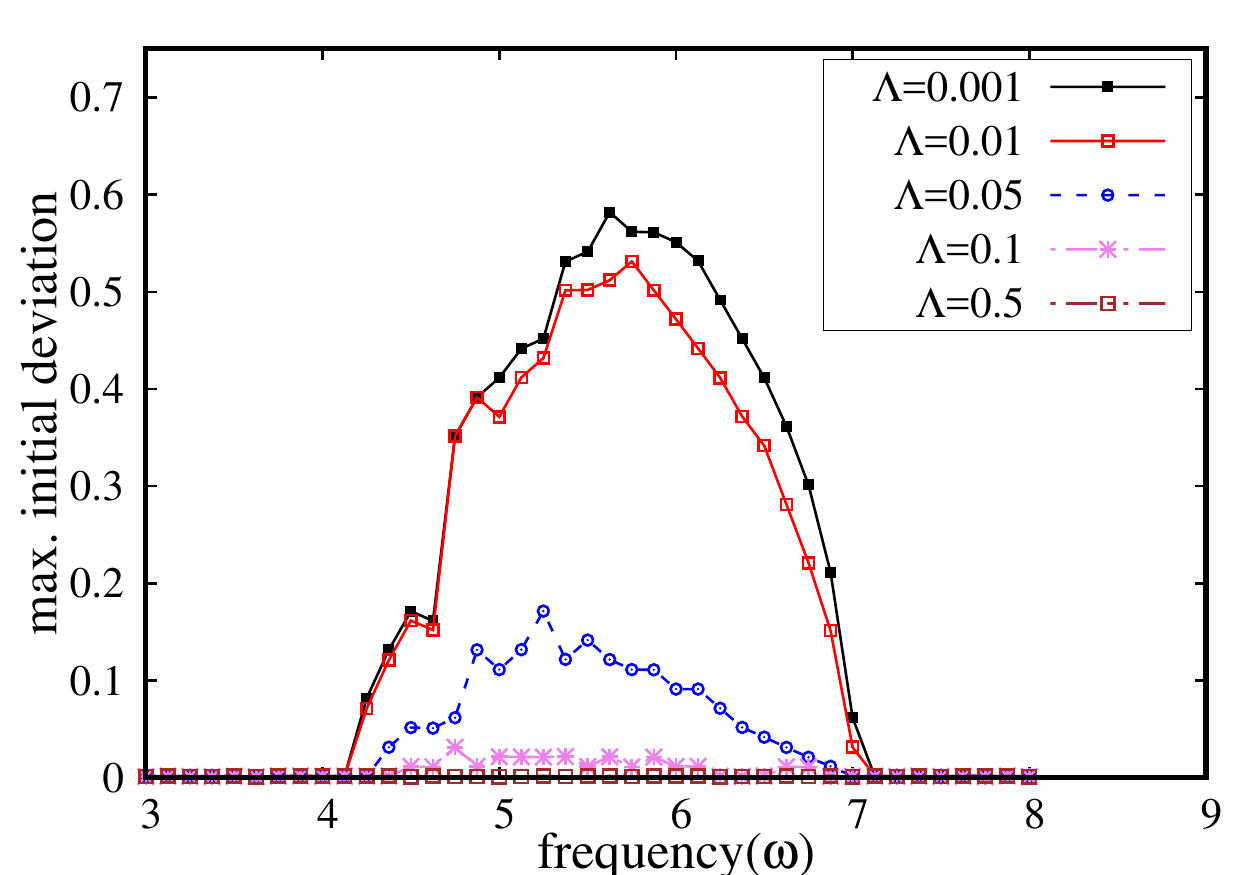}{\footnotesize (b)}\\ \medskip
\includegraphics[width=.45\textwidth]{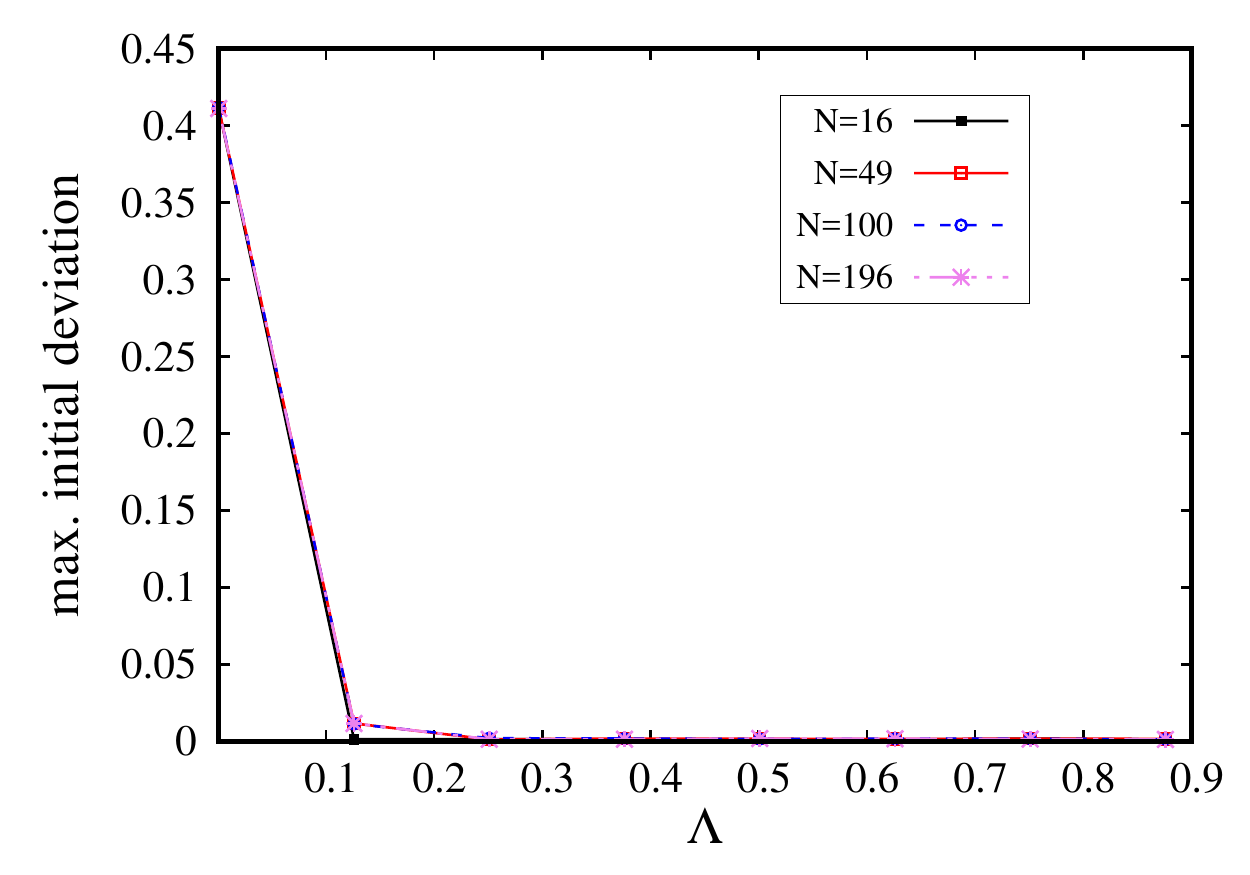}{\footnotesize (c)}
\caption{2D square lattice of Kapitza pendula with nearest neighbour coupling. The  maximum initial deviation from the inverted position $\phi=\pi$, (a) for different values of applied frequency for a set of the coupling term $\Lambda$. Parameters are $g_0=1.0$, $g_1=10$, $N=100$. (b) for different values of applied frequency for a set of the coupling term $\Lambda$. (c)for variation of $\Lambda$ for different sizes of the system. Parameters are $g_0=1.0, \omega=5, N=100$. }
\label{fig:sq_PD}
\end{center}
\end{figure*}

\begin{figure*}[h!]
\centering
\includegraphics[width=1\textwidth]{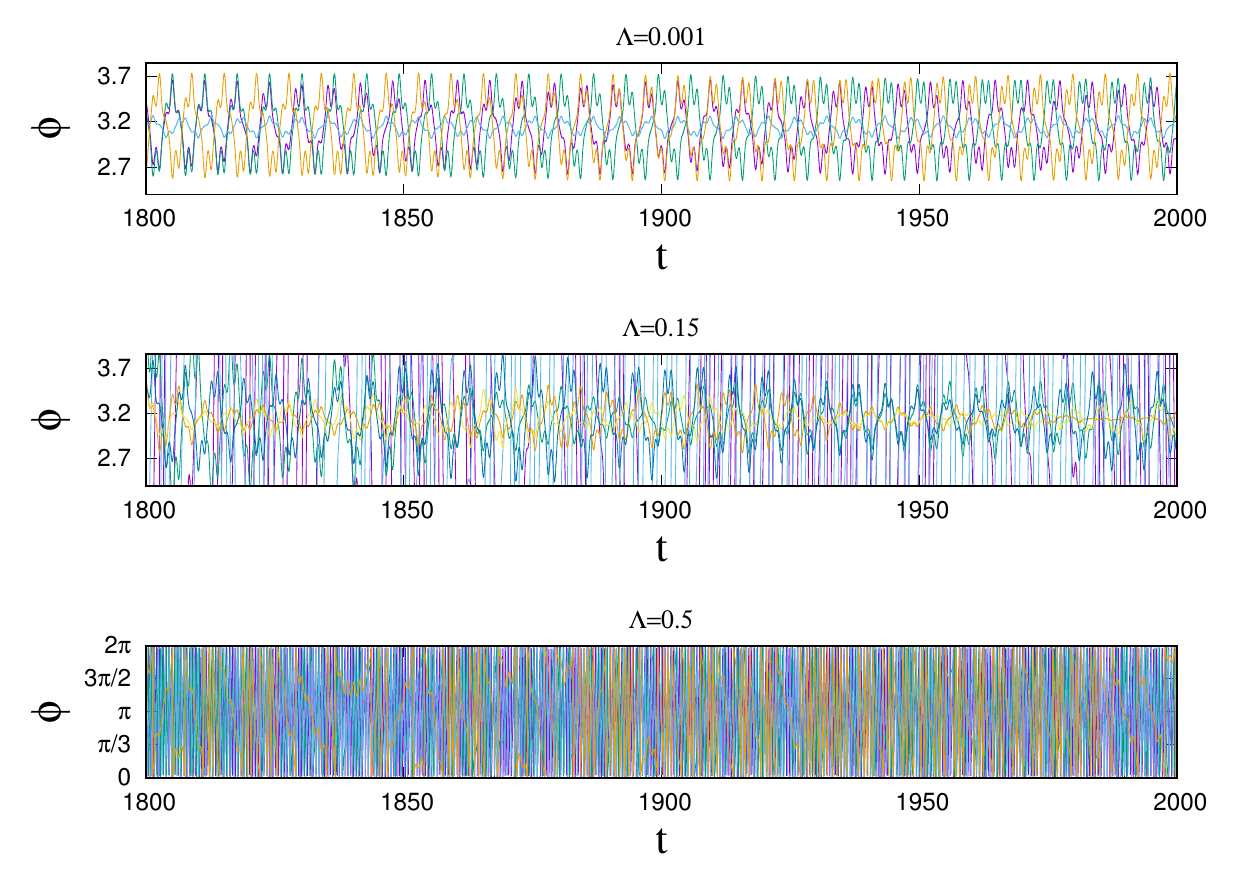}
\caption{Figure shows time-series for three different values of $\Lambda$ for the 2D square network with nearest neighbour coupling. Parameters are $N=1000$. (a) For $\Lambda = 0.001$, all pendula show oscillation arounf $\phi=\pi$. (b) For $\Lambda=0.1$, some oscillates, some rotates. Total integration time $t=2000$. We show time-series for $t=1800$ to $t=2000$ in this figure to remove transients effect.}
\label{fig:ts_sq}
\end{figure*}

\begin{figure}
\centering
\includegraphics[width=.5\textwidth]{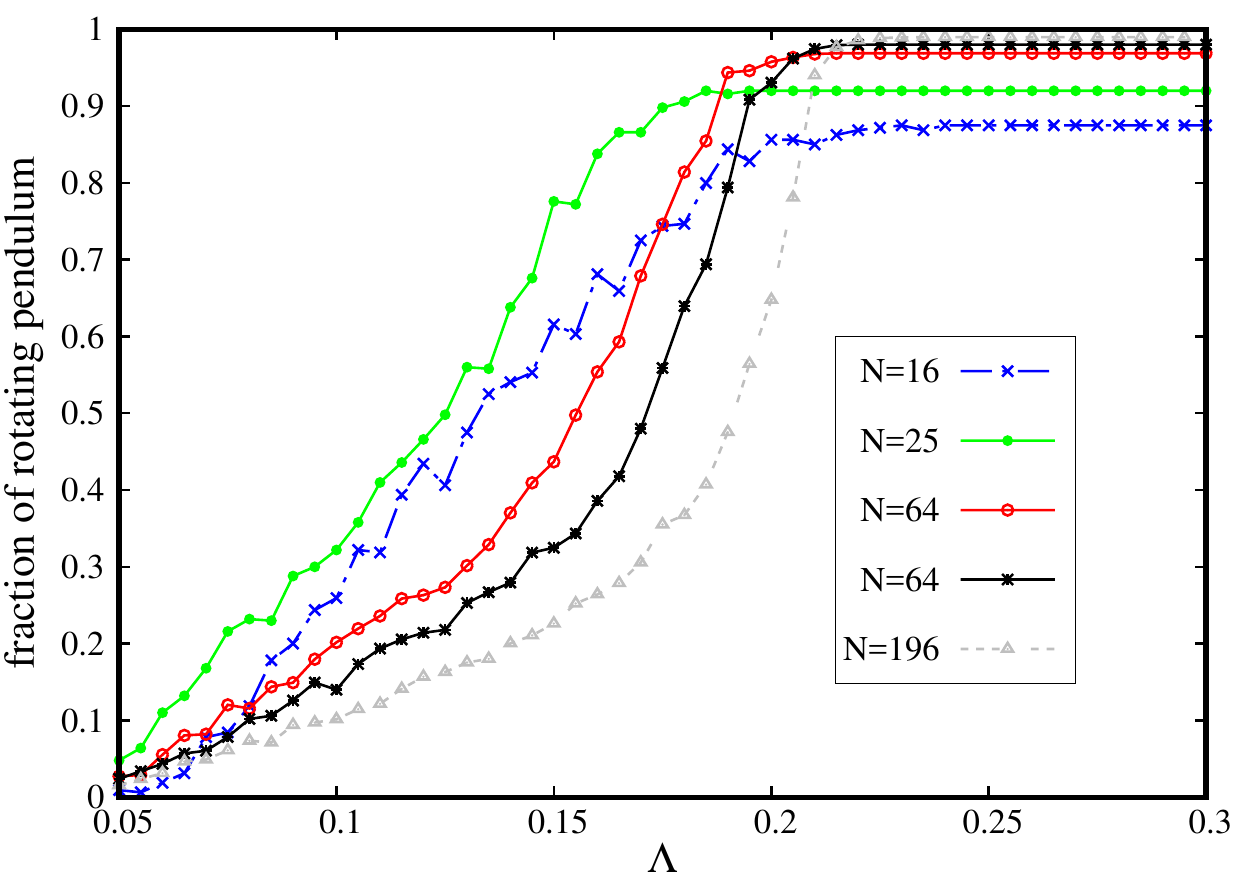}
\caption{Fraction of rotating pendulum for 2D square network. For very small value of $\Lambda$ pendulum shows oscillation around the inverted position. As $\Lambda$ increases some fraction of pendula starts rotating motion. All pendulum starts rotating once the threshold value of $\Lambda$ is passed. }
\label{fig:fracrotsq}
\end{figure}
The numerical results for this system are shown in Fig.\ref{fig:sq_PD}. Fig.\ref{fig:sq_PD}(a) shows the maximum release angle versus $g_1$ curves. As we increase the value of $\Lambda$ the height of
the curve decreases, implying decrement of stability range. However, the rate of decrement is  higher compared to the 1D case. Fig.~\ref{fig:sq_PD}(b) shows the curves for
maximum initial angle versus the driving frequency. Fig.~\ref{fig:sq_PD}(c) shows the effect of the size of the system. We studied the cases for $N=4\times 4, 7\times 7, 10\times 10, 14\times 14$ square lattices. It shows that the stability reduces with increase in $\Lambda$ and the system is insensitive to the system size $N$.
\par

Now we explore the coupled behaviour in the system. The time series is shown in Fig.~\ref{fig:ts_sq}. It shows that for small magnitude of $\Lambda$ all pendula oscillate around  the inverted position. This behavior is lost when $\Lambda$ is increased. For $\Lambda=0.15$, some of the pendula oscillate, whereas, rest of them start rotating as shown in Fig.\ref{fig:ts_sq}(b). As $\Lambda$ is further increased, stability is lost and most of the pendula also go into rotating
motion. In Fig. \ref{fig:fracrotsq}, we show how the fraction of total pendula exhibits oscillation and rotation for different values of $\Lambda$. For very small values, only oscillatory motion around the inverted position is observed. For higher values, all pendula start rotation. There is a range between these two kind of motion where both oscillatory and rotational motion is observed in the system. This is analogous to ``chimera state", simultaneous presence of synchronous(frequency locked) and asynchrounous (rotational motion with no synchrony among the pendula) state in the system.
\subsection{Case 3 : Pendulum chain with all-to-all interaction}
\begin{figure*}
\centering
\includegraphics[width=0.45\textwidth]{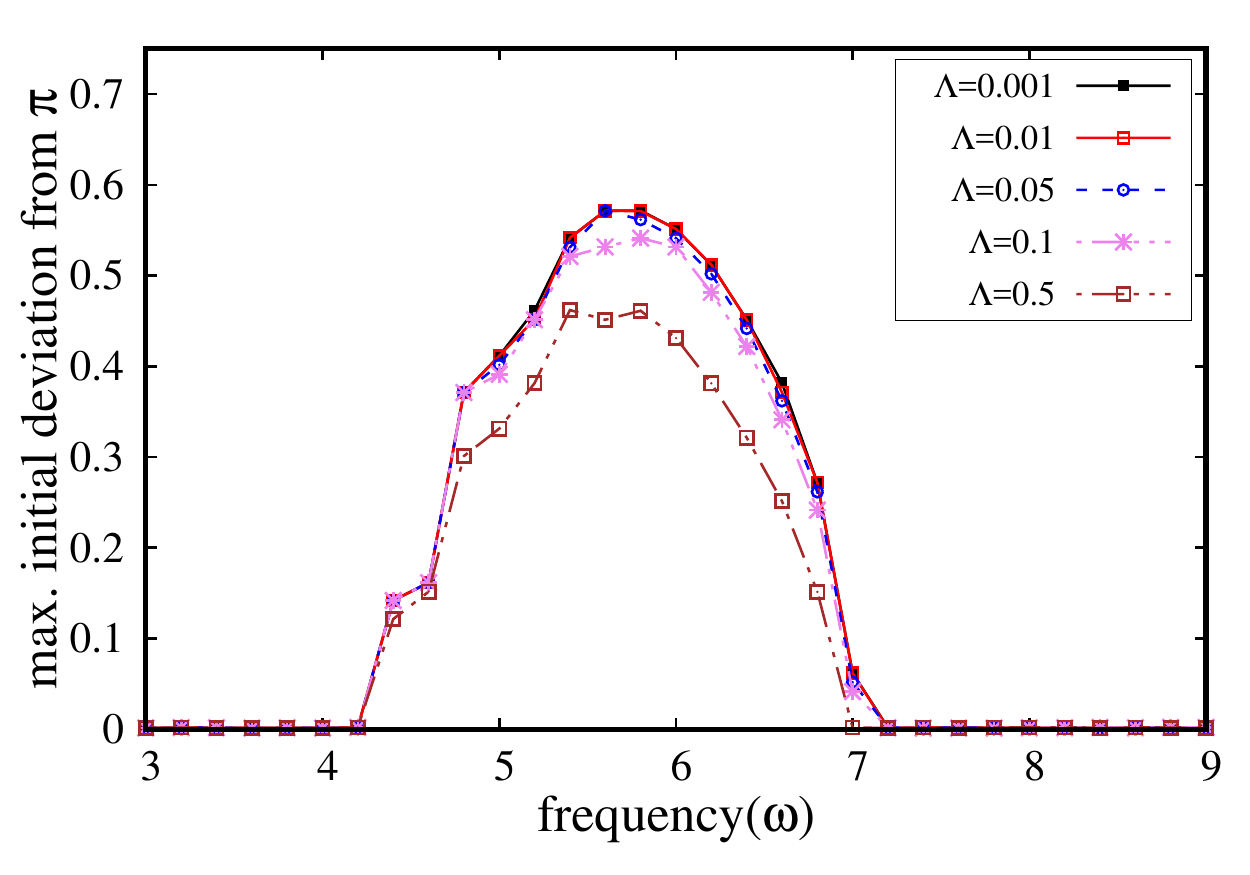}{\footnotesize (a)}\hfill
\includegraphics[width=0.45\textwidth]{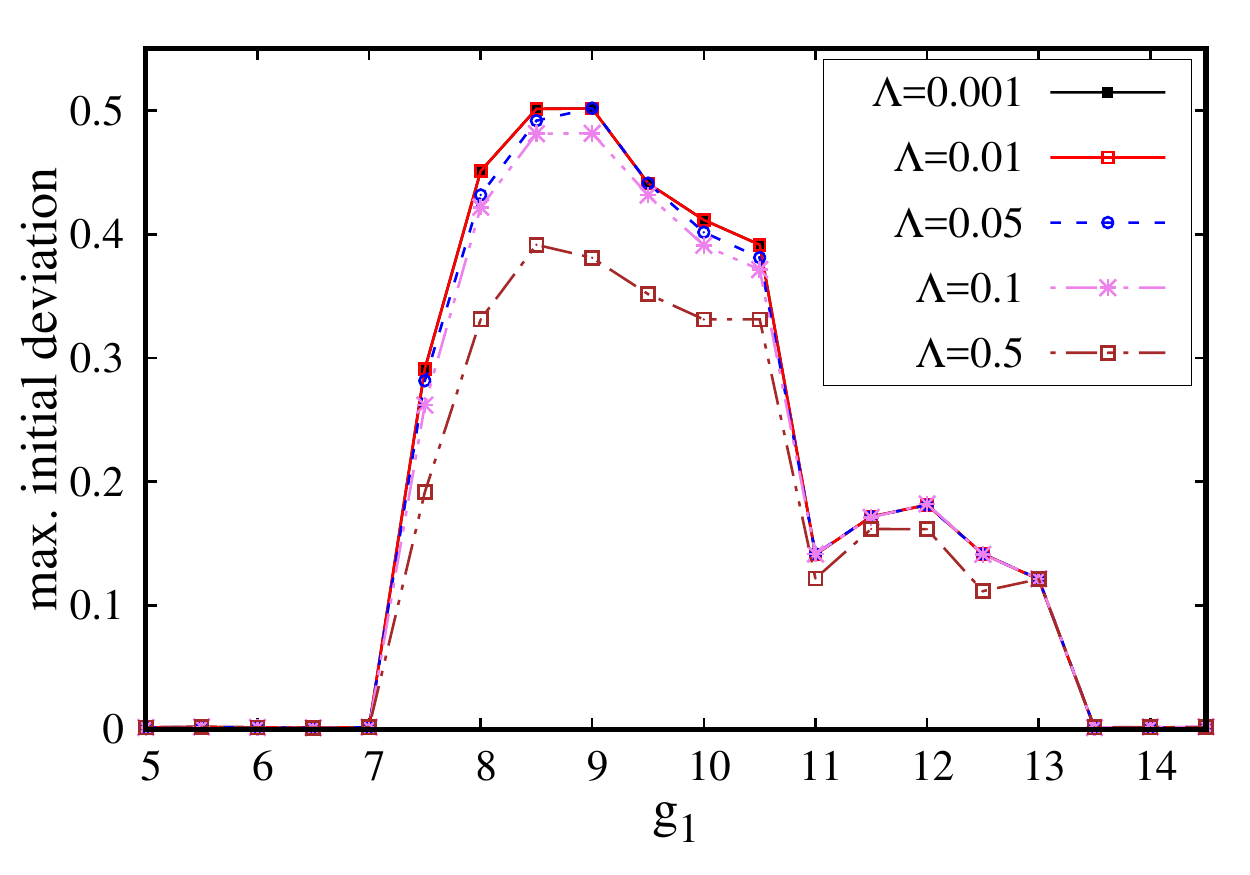}{\footnotesize (b)}\\ \medskip
\includegraphics[width=0.45\textwidth]{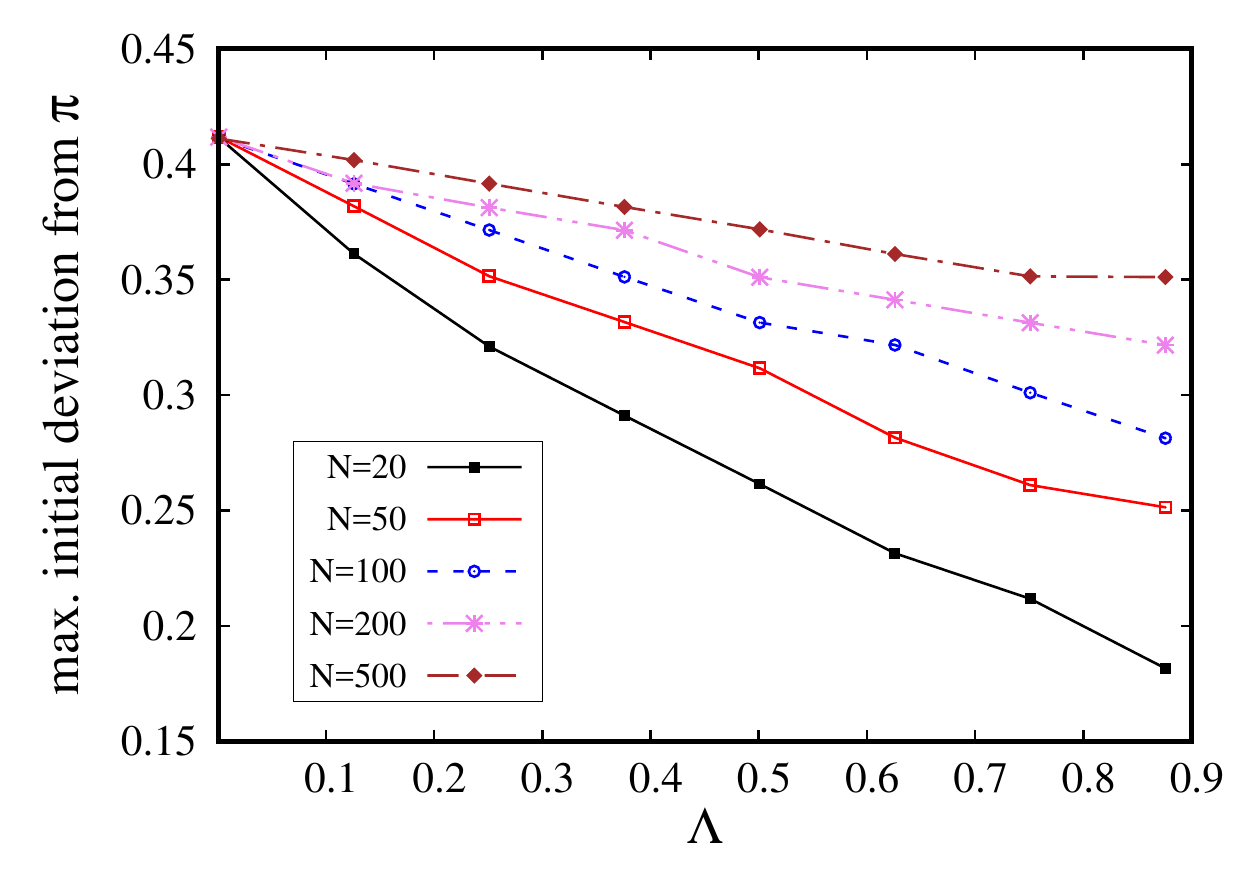}{\footnotesize (c)}
\caption{Inverted pendula with all-to-all coupling. The maximum initial deviation from the inverted position $\phi=\pi$, (a) for different values of applied frequency for a set of coupling term $\Lambda$; the parameters are $g_0=1.0, g_1=10, N=100$, (b) for different values of applied amplitude for a set of the coupling term $\Lambda$; the parameters are $g_0=1.0$, $\omega=5$, $N=100$, (c) for different system sizes ($N$); the parameters are $g_0=1.0,\; g_1=10$ and  $\omega=5$.}
\label{fig:PD_all}
\end{figure*}
In this section we consider many body Kapitza pendulum with all to all interaction. The Hamiltonian governing the dynamics is
\begin{eqnarray}
\label{eq:mbkall}
H=\sum_{i}\Big( \Lambda \frac{p_i^2}{2}-\frac{g(t)}{\Lambda}\cos{\phi_i}\Big)-\frac{\Lambda}{N}\sum_{i,j}\cos(\phi_i-\phi_j)
\end{eqnarray} and equations of motion are 
 \begin{subequations}
  \begin{align}
  \label{eq:alltoalleom}
\dot{\phi_i}&=\Lambda p_i, \\
\dot{p_i}&=-\Big( \frac{g(t)}{\Lambda}\sin{\phi_i}+\frac{\Lambda}{N}\sin(\phi_i-\phi_j)\Big).
 \end{align}
 \end{subequations}
 The interaction term is scaled by the total number of pendula $N$ to make it thermodynamically stable. Applying mean field theory we can define an order parameter for such a system with long range interaction \cite{Antoni1995}. The order parameter for the system is of the form 
\begin{eqnarray}
\label{eq:coherence}
\mathbf{R}=r e^{i\phi}=\frac{1}{N}\sum_{i=i}^N \mathbf{r}_i
\end{eqnarray}
where $r$ and $\phi$ represent the modulus and the phase of the order parameter and $\mathbf{r}_i=(\cos{\phi_i},\sin{\phi_i})$. Now the potential can be rewritten as a sum of single particle potentials 
\begin{eqnarray}
v_i=1-\Lambda M\cos(\theta_i-\phi).
\end{eqnarray}
   Fig.\ref{fig:PD_all}(a) shows the maximum release angle versus $g_1$ curves for this system. With the increase of $\Lambda$, the height of the curve decreases which implies the decrement of the stability range. Compared to the 1D and 2D case, the rate of decrement is least for this all-to-all coupling. Fig.~\ref{fig:PD_all}(b) shows the curves for
maximum initial angle versus the driving frequency. To observe the effect of system size we studied the cases for $N = 20, 50, 100, 200, 500$. Fig.~\ref{fig:PD_all}(c) shows that for a fixed value of $\Lambda$ the stability increases as we increase the size of the system.
\par
\begin{figure}[h]
\centering
\includegraphics[width=1\textwidth]{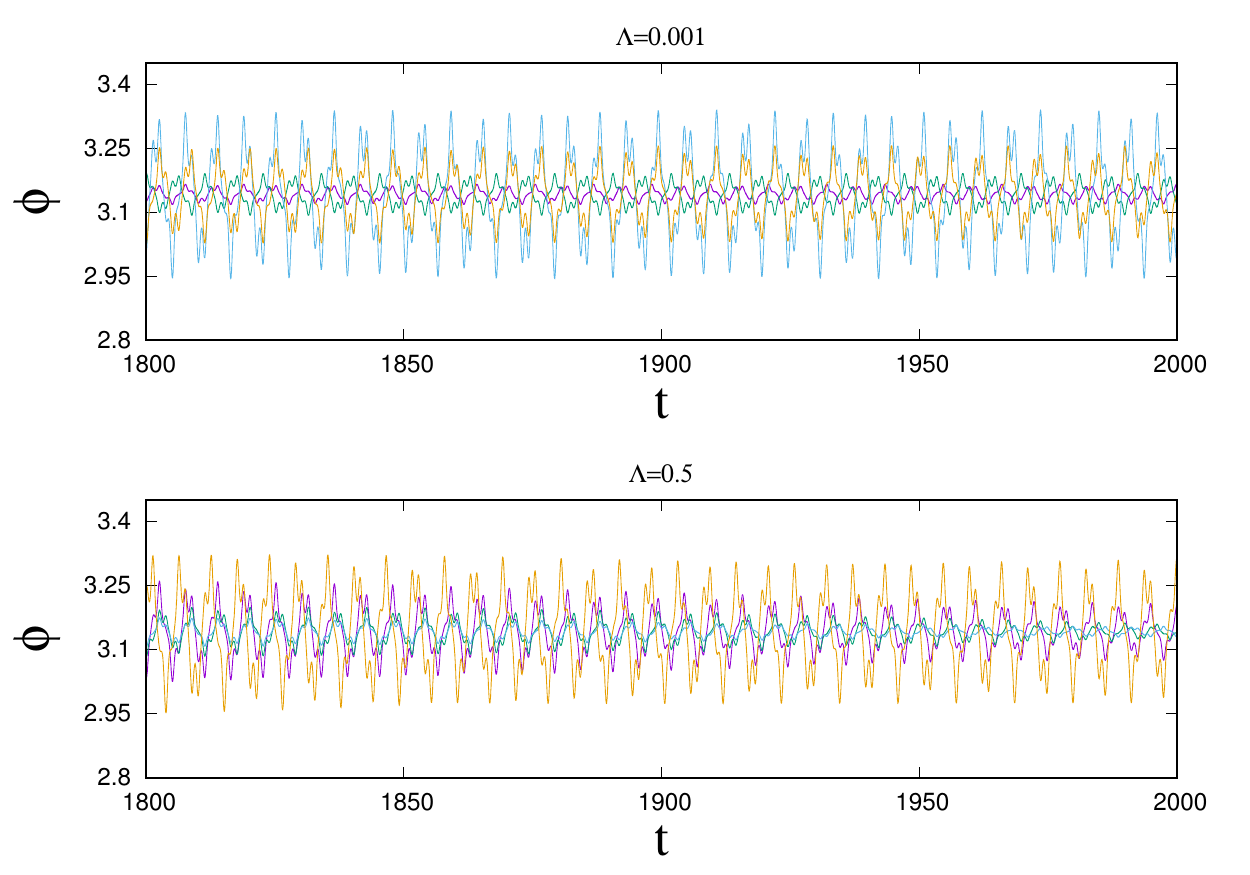}
\caption{Time series for all to all coupling for $\Lambda=0.001$(upper panel) and $\Lambda=0.5$ (lower panel). Parameters are : $N=1000, g_0=1, g_1= 10, \omega=5$. In both cases, all pendula oscillate around $\phi=\pi$ (we have shown only $5$ pendula in this figure). Total integration time $t_f=2000$ and to avoid transient effect we show $t=1800$ to $t=2000$ in this figure.}
\label{fig:ts_all}
\end{figure}
\begin{figure}[h!]
\centering
\includegraphics[width=0.85\textwidth]{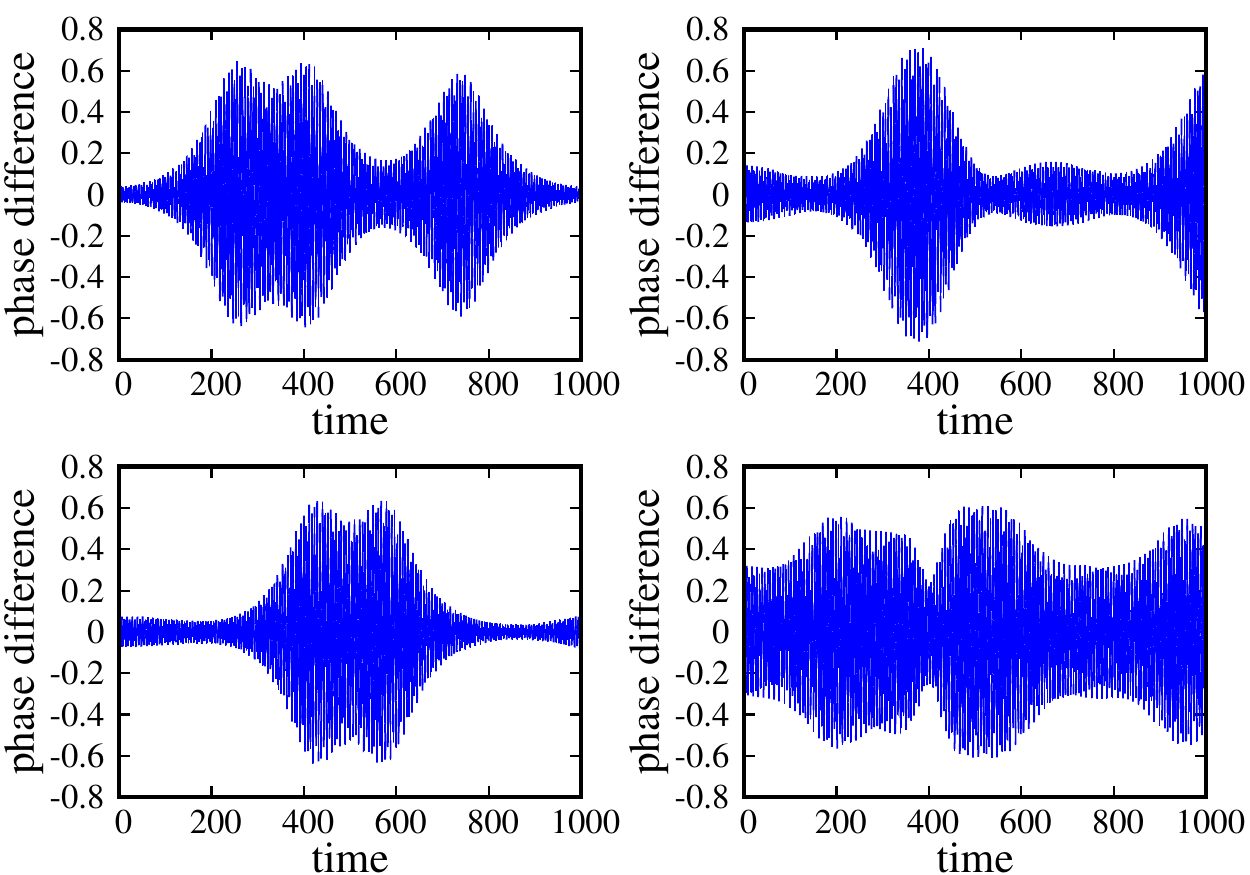}
\caption{ This figure shows the time-series of relative angular displacement $\delta \phi=\phi_i-\phi_1$, where indices $1$ and $i$ corresponds to $1\textsuperscript{st}$ and $i\textsuperscript{th}$ pendulum. This is when $\Lambda=1$. These specific envelops correspond to Group - \Rmnum{2}, \Rmnum{4}, \Rmnum{5} and Group -\Rmnum{6} of Table. \ref{table:4}.}
\label{fig:envelop}
\end{figure}

We show the time series in Fig.~\ref{fig:ts_all} for $\Lambda=0.001$ and $\Lambda=0.5$. Unlike 1D and 2D case, we observe only oscillation around the inverted position in this range of $\Lambda$; we do not see any rotation in all-to-all coupling. However, in this case, beat appears in the system. At low values of $\Lambda$ clustering is not apparent. But careful examination reveals that the beats observed in the individual pendula are not the same. A few examples of different beat patterns are shown in Fig. \ref{fig:envelop}. A few pendula form  groups that follow the same envelop of the beat pattern.
\begin{center}
\begin{table}[ht]
\begin{minipage}[b]{0.45\linewidth}\centering
  \begin{tabular}{ |p{2cm}||p{2.5cm}| }
 \hline
  Group & No.of pendula \\ [0.5ex] 
 \hline\hline
1&11\\
2&6\\
3&7\\
4&4\\
5&9\\
6&5\\
7&2\\
8&2\\
\hline
not clustered & 2\\  
  [1ex] 
 \hline
\end{tabular}
\caption{Table showing clustered and unclustered pendulum for $\Lambda
=0.001.$}
\label{table:1}
\end{minipage}
\hspace{0.2cm}
\begin{minipage}[b]{0.45\linewidth}
  \begin{tabular}{ |p{2cm}||p{2.5cm}| }
 \hline
  Group & No.of pendula \\ [0.5ex] 
 \hline\hline
1&13\\
2&3\\
3&5\\
4&10\\
5&4\\
6&3\\
7&2\\
8&2\\
9&4\\
\hline
not clustered & 3\\  
  [1ex] 
 \hline
\end{tabular}
\caption{Table showing clustered and unclustered pendulum for $\Lambda
=0.1.$}
\label{table:2}
\end{minipage}
\end{table}
\end{center}

\begin{table}
\begin{minipage}[b]{0.45\linewidth}
 \begin{tabular}{ |p{2cm}||p{2.5cm}| }
 \hline
  Group & No.of pendula \\ [0.5ex] 
 \hline\hline
1&5\\
2&5\\
3&3\\
4&4\\
5&13\\
6&8\\
7&2\\
8&2\\
9&2\\
10&3\\
\hline
not clustered & 2\\  
  [1ex] 
 \hline
\end{tabular}

\caption{Table shows groups of clustered pendula. We consider $N=50, \Lambda=0.5$.}
\label{table:3}
\end{minipage}
\hspace{0.3cm}
\begin{minipage}[b]{0.45\linewidth}
 \begin{tabular}{ |p{2cm}||p{2.5cm}| }
 \hline
  Group & No.of pendula \\ [0.5ex] 
 \hline\hline
1&13\\
2&13\\
3&9\\
4&5\\
5&5\\
6&2\\
\hline
not clustered & 2\\  
  [1ex] 
 \hline
\end{tabular}
\caption{Table shows the number of pendula belonging to different clusters when $\Lambda=1, N=50$.}
\label{table:4}
\end{minipage}

\end{table}

 We consider $N=50$ and catagorise the pendula into group according to the envelop they belong to. List of pendula and their corresponding group is presented in Table \ref{table:1}, \ref{table:2}, \ref{table:3} and \ref{table:4}. As soon as $\Lambda$ increases pendula start to form clusters. Some of them remain unclustered. In the clustered group, some pendula may have some phase difference. However, as we increase $\Lambda$, the phase difference vanishes and pendula are in complete synchrony with each other for a specific envelop. 
 
Our intention was to show that clustering exists. However, which specific pendulum will belong to which cluster may vary from simulation to simulation depending on the choice of (random) initial conditions. But the number of pendula in a  group remains more or less same.

\section{Conclusions}\label{sec:Discussion}
In this paper we have considered systems of coupled Kapitza pendula in
three different network structures. We have studied the effects of the coupling between pendula (embodied in the coupling parameter $\Lambda$ in our formulation), the network dimension ($D$), and the system size ($N$) on the stability of the system.

Increase in the coupling parameter $\Lambda$ is
found to reduce the stability margins. The effect of $\Lambda$ is significant in 1D lattice but becomes very pronounced in 2D lattice with nn coupling. The effect, however, reduces for all-to-all coupling. The system size, i.e., the number of coupled pendula $N$, also plays
an important role. For the  all-to-all coupled system, an increase
in system size drastically improves the stability of the system. But
the nearest neighbour coupled lattice is relatively insensitive to
the system size.\par
We have also explored the collective behaviour in these three networks for $\Lambda \sim [0.01, 1.0]$. As we increase the value of $\Lambda$, rotational motion appears in nn coupled 1D and 2D networks. However, we do not see any such rotation in the all-to-all coupling but in this case beat appears in the system. Careful study of the beating in the all-to all case, reveals the appearance of several shapes of envelop. Once the individual time-series are catagorised into different groups corresponding to the different shapes of the envelops, clustering becomes evident.
\begin{acknowledgements}
The author would like to thank Soumitro Banerjee and Emanuele G. Dalla Torre for valuable suggestions and comments on the manuscript. 
\end{acknowledgements}


\begin{thebibliography}{10}

\bibitem{Paul1990}
Wolfgang Paul.
\newblock Electromagnetic traps for charged and neutral particles.
\newblock {\em Rev. Mod. Phys.}, 62:531--540, Jul 1990.

\bibitem{Gilary2003}
Ido Gilary, Nimrod Moiseyev, Saar Rahav, and Shmuel Fishman.
\newblock Trapping of particles by lasers: the quantum kapitza pendulum.
\newblock {\em Journal of Physics A: Mathematical and General}, 36(25):L409,
  2003.

\bibitem{Bullo2002}
Francesco Bullo.
\newblock Averaging and vibrational control of mechanical systems.
\newblock {\em SIAM Journal on Control and Optimization}, 41(2):542--562, 2002.

\bibitem{Nakamura1997}
Y.~Nakamura, T.~Suzuki, and M.~Koinuma.
\newblock Nonlinear behavior and control of a nonholonomic free-joint
  manipulator.
\newblock {\em IEEE Transactions on Robotics and Automation}, 13(6):853--862,
  Dec 1997.

\bibitem{Stephenson1908}
Andrew Stephenson.
\newblock XX. on induced stability.
\newblock {\em Philosophical Magazine Series 6}, 15(86):233--236, 1908.

\bibitem{Haar(Eds.)1965}
D.~ter Haar~(Eds.).
\newblock {\em Collected Papers of P.L. Kapitza. Volume 2}.
\newblock pergamon, 1st edition, 1965.

\bibitem{Kim1998}
Sang-Yoon Kim and Bambi Hu.
\newblock Bifurcations and transitions to chaos in an inverted pendulum.
\newblock {\em Phys. Rev. E}, 58:3028--3035, Sep 1998.

\bibitem{Butikov2001}
Eugene~I. Butikov.
\newblock On the dynamic stabilization of an inverted pendulum.
\newblock {\em American Journal of Physics}, 69(7):755--768, 2001.

\bibitem{Broer1999}
H.W. Broer, I.~Hoveijn, M.~van Noort, and G.~Vegter.
\newblock The inverted pendulum: A singularity theory approach.
\newblock {\em Journal of Differential Equations}, 157(1):120 -- 149, 1999.

\bibitem{Broer2004}
H.~W. Broer, I.~Hoveijn, M.~van Noort, C.~Sim{\'o}, and G.~Vegter.
\newblock The parametrically forced pendulum: A case study in 1 1/2 degree of
  freedom.
\newblock {\em Journal of Dynamics and Differential Equations}, 16(4):897--947,
  2004.

\bibitem{Bartuccelli2001}
M.~V. Bartuccelli, G.~Gentile, and K.~V. Georgiou.
\newblock On the dynamics of a vertically driven damped planar pendulum.
\newblock {\em Proceedings of the Royal Society of London A: Mathematical,
  Physical and Engineering Sciences}, 457(2016):3007--3022, 2001.

\bibitem{Bartuccelli2002}
M.~V. Bartuccelli, G.~Gentile, and K.~V. Georgiou.
\newblock On the stability of the upside{\textendash}down pendulum with
  damping.
\newblock {\em Proceedings of the Royal Society of London A: Mathematical,
  Physical and Engineering Sciences}, 458(2018):255--269, 2002.

\bibitem{Bartuccelli2003}
Michele~V Bartuccelli, Guido Gentile, and Kyriakos~V Georgiou.
\newblock Kam theory, lindstedt series and the stability of the upside-down
  pendulum.
\newblock {\em Discrete and Continuous Dynamical Systems}, 9(2):413--426, 2003.

\bibitem{Eckardt2017}
Andr\'e Eckardt.
\newblock Colloquium.
\newblock {\em Rev. Mod. Phys.}, 89:011004, Mar 2017.

\bibitem{Chacon2010}
R~Chacon and L~Marcheggiani.
\newblock Controlling spatiotemporal chaos in chains of dissipative kapitza
  pendula.
\newblock {\em Physical Review E}, 82(1):016201, 2010.

\bibitem{Marcheggiani2014}
Laura Marcheggiani, Ricardo Chac{\'o}n, and S~Lenci.
\newblock On the synchronization of chains of nonlinear pendula connected by
  linear springs.
\newblock {\em The European Physical Journal: Special Topics}, 223(4):729--756,
  2014.

\bibitem{Moehlis2008}
Jeff Moehlis.
\newblock On the dynamics of coupled parametrically forced oscillators.
\newblock In {\em Proceedings of 2008 ASME Dynamic Systems and Control
  Conference. Ann Arbor, Michigan, USA}, 2008.

\bibitem{Danzl2010}
Per Danzl and Jeff Moehlis.
\newblock Weakly coupled parametrically forced oscillator networks: existence,
  stability, and symmetry of solutions.
\newblock {\em Nonlinear Dynamics}, 59(4):661--680, Mar 2010.

\bibitem{Xu2014}
Y.~Xu, T.~J. Alexander, H.~Sidhu, and P.~G. Kevrekidis.
\newblock Instability dynamics and breather formation in a horizontally shaken
  pendulum chain.
\newblock {\em Phys. Rev. E}, 90:042921, Oct 2014.

\bibitem{SalgadoSanchez2016}
P.~Salgado~S\'anchez, J.~Porter, I.~Tinao, and A.~Laver\'on-Simavilla.
\newblock Dynamics of weakly coupled parametrically forced oscillators.
\newblock {\em Phys. Rev. E}, 94:022216, Aug 2016.

\bibitem{Citro2015}
Roberta Citro, Emanuele~G Dalla~Torre, Luca D’Alessio, Anatoli Polkovnikov,
  Mehrtash Babadi, Takashi Oka, and Eugene Demler.
\newblock Dynamical stability of a many-body kapitza pendulum.
\newblock {\em Annals of Physics}, 360:694--710, 2015.

\bibitem{Eckardt2005}
Andr\'e Eckardt, Christoph Weiss, and Martin Holthaus.
\newblock Superfluid-insulator transition in a periodically driven optical
  lattice.
\newblock {\em Phys. Rev. Lett.}, 95:260404, Dec 2005.

\bibitem{Lignier2007}
H.~Lignier, C.~Sias, D.~Ciampini, Y.~Singh, A.~Zenesini, O.~Morsch, and
  E.~Arimondo.
\newblock Dynamical control of matter-wave tunneling in periodic potentials.
\newblock {\em Phys. Rev. Lett.}, 99:220403, Nov 2007.

\bibitem{Struck2011}
J.~Struck, C.~{\"O}lschl{\"a}ger, R.~Le~Targat, P.~Soltan-Panahi, A.~Eckardt,
  M.~Lewenstein, P.~Windpassinger, and K.~Sengstock.
\newblock Quantum simulation of frustrated classical magnetism in triangular
  optical lattices.
\newblock {\em Science}, 333(6045):996--999, 2011.

\bibitem{Struck2012}
J.~Struck, C.~\"Olschl\"ager, M.~Weinberg, P.~Hauke, J.~Simonet, A.~Eckardt,
  M.~Lewenstein, K.~Sengstock, and P.~Windpassinger.
\newblock Tunable gauge potential for neutral and spinless particles in driven
  optical lattices.
\newblock {\em Phys. Rev. Lett.}, 108:225304, May 2012.

\bibitem{Russomanno2012}
Angelo Russomanno, Alessandro Silva, and Giuseppe~E. Santoro.
\newblock Periodic steady regime and interference in a periodically driven
  quantum system.
\newblock {\em Phys. Rev. Lett.}, 109:257201, Dec 2012.

\bibitem{Ponte2015}
Pedro Ponte, Anushya Chandran, Z.~Papić, and Dmitry~A. Abanin.
\newblock Periodically driven ergodic and many-body localized quantum systems.
\newblock {\em Annals of Physics}, 353:196 -- 204, 2015.

\bibitem{Ponte2015a}
Pedro Ponte, Z.~Papi\ifmmode~\acute{c}\else \'{c}\fi{}, Fran\ifmmode
  \mbox{\c{c}}\else~\c{c}\fi{}ois Huveneers, and Dmitry~A. Abanin.
\newblock Many-body localization in periodically driven systems.
\newblock {\em Phys. Rev. Lett.}, 114:140401, Apr 2015.

\bibitem{Blackburn1992}
James~A. Blackburn, H.~J.~T. Smith, and N.~Gro/nbech‐Jensen.
\newblock Stability and hopf bifurcations in an inverted pendulum.
\newblock {\em American Journal of Physics}, 60(10):903--908, 1992.

\bibitem{Antoni1995}
Mickael Antoni and Stefano Ruffo.
\newblock Clustering and relaxation in hamiltonian long-range dynamics.
\newblock {\em Phys. Rev. E}, 52:2361--2374, Sep 1995.

\end{thebibliography}
\end{document}